\documentclass[
twocolumn,
tightenlines,
10pt,
longbibliography,
showpacs,
nofootinbib,
notitlepage,
superscriptaddress]{revtex4-2}

\usepackage{amsmath, amssymb, amsthm, braket, enumitem, dsfont, relsize}
\usepackage{makecell, graphicx, subfiles, multirow, wrapfig, tabularx}
\usepackage[font=scriptsize]{subcaption}
\usepackage[font=scriptsize, justification=raggedright]{caption}
\usepackage[page]{appendix}
\usepackage[colorlinks, allcolors=black]{hyperref}
\usepackage[capitalize, nameinlink]{cleveref}
\usepackage{tikz}
\usetikzlibrary{quantikz}
\usetikzlibrary{shapes.geometric, arrows}
\usetikzlibrary{positioning}
\setlist[itemize]{leftmargin=10pt}

\DeclareCaptionLabelFormat{bold}{\textbf{(#2)}}
\begin{document}

\title{Non-convex Quadratic Programming Using Coherent Optical Networks}

\author{Farhad~Khosravi}
\affiliation{1QB Information Technologies (1QBit), Vancouver, BC, Canada}

\author{Ugur~Yildiz}
\affiliation{1QB Information Technologies (1QBit), Vancouver, BC, Canada}

\author{Artur~Scherer}
\affiliation{1QB Information Technologies (1QBit), Vancouver, BC, Canada}

\author{Pooya~Ronagh}
\thanks{{\vskip-10pt}{\hskip-8pt}Corresponding author: \href{mailto:pooya.ronagh@1qbit.com}{pooya.ronagh@1qbit.com}\\}
\affiliation{1QB Information Technologies (1QBit), Vancouver, BC, Canada}
\affiliation{Institute for Quantum Computing, University of Waterloo, Waterloo, ON, Canada}
\affiliation{Department of Physics \& Astronomy, University of Waterloo, Waterloo, ON, Canada}
\affiliation{Perimeter Institute for Theoretical Physics, Waterloo, ON, Canada}

\date{\today}

\begin{abstract}
We investigate the possibility of solving continuous non-convex optimization
problems using a network of interacting quantum optical oscillators. We propose a native encoding of continuous variables in analog signals associated with the 
quadrature operators of a set of quantum optical modes. Optical coupling of the modes and noise introduced 
by vacuum fluctuations from external reservoirs or by weak measurements of the modes are used to optically simulate a diffusion process on a set of continuous random variables.
The process is run sufficiently long for it to relax into the steady state
of an energy potential defined on a continuous domain. As a first
demonstration, we numerically benchmark solving box-constrained quadratic
programming (BoxQP) problems using these settings. We consider delay-line 
 and measurement-feedback variants of the experiment. Our benchmarking
results demonstrate that in both cases the optical network is capable of solving BoxQP
problems over three orders of magnitude faster than a state-of-the-art classical
heuristic.
\end{abstract}

\maketitle

\section{Introduction}

Real-world optimization problems frequently involve decision variables that are
naturally continuous. This is the case for continuous non-convex global
optimization problems and the more general class of mixed-integer programming
(MIP) models~\cite{burer2012nonconvex} designed to solve a variety of NP-hard
optimization problems of high practical relevance, including, for example,
scheduling problems, flow problems, resource allocation problems, and routing
problems. Even the special class of quadratic non-convex global optimization
problems captures extremely challenging computational tasks that are
NP-hard~\cite{pardalos1988checking, burer2009globally}. Classical
optimization techniques lack satisfactory performance in solving
continuous non-convex optimization problems at scale~\cite{burer2009globally,
bonami2018globally, vries2022tight}.

In recent years, the design of special-purpose computing architectures for solving optimization problems has been gaining increasing attention~\cite{mohseni2022ising}. Common discrete optimization engines include
physics-inspired heuristics such as thermal (i.e., simulated)
annealing~\cite{kirkpatric1983optimization, johnson1989optimization} and its
variants~\cite{aramon2019physics, takemoto20192}, quantum
annealing~\cite{defalco1988numerical, apolloni1989quantum, kadowaki1998quantum,
brooke1999quantum}, coherent Ising machines (CIM)~\cite{wang2013coherent,
yamamoto2017coherent}, local heuristic search techniques~\cite{benlic2017breakout, prajapati2020tabu}, and various spin-based Monte Carlo
methods. All these approaches focus on solving discrete optimization problems
natively, and even in that class, the attention has been limited to solving Ising
problems. This means that more-general types of problems (e.g., those
comprising continuous variables) require processing at the software level.
Algorithmic problem reformulation and reduction is not only costly in terms of
the computing resources it requires, but it can result in ill-behaved energy
landscapes that are difficult to optimize. Such schemes hinder the
practical applicability of the resultant technologies, as evidenced by over a
decade's worth of unsuccessful attempts to find practical applications for the technologies.

As demonstrated by several previous studies~\cite{hamerly2019experimental,
leleu2019destabilization, leleu2021scaling, reifenstein2021coherent,
sankar2021benchmark}, coherent-network computing has appeared as a promising
approach for solving NP-hard combinatorial optimization problems. The well-known time-multiplexed CIMs utilize a network of degenerate optical parametric
oscillators (degenerate OPO or DOPO) that are injected into a ring cavity and
gradually pumped at a rate well above the bifurcation
threshold~\cite{wang2013coherent, yamamoto2017coherent}. These CIMs can be realized through either fully optical experiments involving optical delay
lines (DL-CIM)~\cite{takata2015quantum, maruo2016truncated}, or using the assistance of a digital processing device (e.g., an FPGA) in a
measurement-feedback procedure (MF-CIM) ~\cite{leleu2019destabilization,
kako2020coherent}. While the digital processor
may limit the performance of the measurement-feedback variant, in principle, by
operating at optical frequencies, a CIM operates at a much faster clock speed
and far lower power consumption compared to digital computers. With optical
pulses that are only a few picoseconds apart, a CIM can solve binary
optimization problems involving thousands of variables in a fraction of a
second~\cite{mcmahon2016fully}.

Pumping DOPOs above their bifurcation threshold forces their amplitudes to
attain certain extreme positive or negative values. In this paper, we show that if the
DOPOs are pumped less aggressively, their mean-field amplitudes can 
converge to stable fractional values. When solving binary optimization problems,
conventional CIMs begin an optimization process as quantum analog devices and
finish it as classical digital devices~\cite{yamamoto2020coherent}. In the case of a device that exhibits very low signal loss, a product of coherent cat states of the form $|\alpha \rangle + |-\alpha \rangle$ is generated at slightly above the bifurcation threshold~\cite{kiesewetter2022coherent}. Iterative weak measurements then result in a collapse into a product of classical bits representing a low-energy state of the optimization problem. While this quantum-to-classical transition is widely considered to be a key characteristic of the CIM~\cite{yamamoto2020coherent}, a clear computational mechanism for why low-energy states are favoured by this process is not apparent. 

In our work, we focus on the Langevin dynamics in the CIM, as opposed to the quantum-to-classical transition, as the mechanism responsible for solving optimization problems, with the diffusion process generated by the iterative noise injection using the nonlinear crystal and the measurement process. 
Langevin dynamics is known to mix into low-energy states of the dynamical system's potential~\cite{chiang1987diffusion,xu2017global}. However, in the CIM, the drift term deviates from the gradient of the objective function. This discrepancy, along with the non-convex landscape
of the optimization problem, prevent the CIM from always succeeding in finding the global optimum. In order to test the hypothesis that Langevin dynamics in the CIM is the mechanism responsible for solving optimization problems, we demonstrate its effectiveness in solving non-convex continuous optimization problems without the need to prepare cat states. Indeed, being intrinsically continuous, the analog amplitudes of
the DOPO pulses can be used to represent continuous variables of a continuous
optimization problem below the saturation threshold. In order to encourage the realization of
experiments beyond Ising optimization, we refer to our setup as a coherent
continuous-variable machine (CCVM).

At the pump rates at which a CCVM is operated, the entire time evolution is carried out
in the coherent regime of the device. While this itself opens the door to
investigating the signatures of a quantum parallel search, we do not focus on such
aspects of the operation of the machine, and instead view the machine as an
optical device intended to \emph{optically} perform a Brownian motion process which is 
then used to \emph{optically} integrate a diffusion process of the form
\begin{equation}
dc= b(t, c)\, dt -\nabla f(c)\, dt + \sigma(t)\, dW.
\end{equation}
Here, $c$ represents the mean amplitude and phase of the DOPO pulses, $f$ is a
continuous function we are interested in minimizing, and $b(t, c)$ is a
component-wise force driving the dynamics of each DOPO, which are controlled by
the DOPO's pump rate. The variable $W$ is a standard Wiener process realized by weak
measurements of the optical modes either via photon loss in a highly
dissipative cavity, particularly in a delay-line CCVM (\mbox{DL-CCVM}), or via
out-coupling and homodyne detection in a measurement-feedback CCVM
(MF-CCVM). Finally, $\sigma (t)$ represents the diffusion rate, and is
controlled via the usage of cavities with a certain amount of finesse in the DL-CCVM, or
by controlling the out-coupling rate of beamsplitters in the MF-CCVM.

Monte Carlo simulations of such stochastic differential equations (SDE) have provided a
breath of techniques in digital computing for solving global optimization
problems~\cite{xu2017global, raginsky2017non}. Indeed, the convergence
rate of diffusion processes of the form presented above are well-studied. For instance, 
if the drift is a conservative force satisfying Poincar{\'e} or log-Sobolov
inequalities, then the process mixes exponentially fast into the Gibbs distribution
of the potential, but for highly non-convex potentials the mixing time is
exponentially long. We do not propose employing coherent computing as a way to avoid
exponentially long mixing times. Instead, we are interested in proposing power-efficient computing architectures that \emph{naturally} integrate
SDEs of the form given above by exploiting the quantum mechanical properties of light.

Our long-term goal is to explore the capabilities of CCVMs in solving continuous and
mixed-integer non-convex optimization problems natively, thereby avoiding the
aforementioned overhead. As a first step, we benchmark the performance of our
proposed technology by predicting the time-to-solution (TTS) scaling of CCVMs for
solving a simple example of continuous non-convex optimization problems, that is,
the {\em box-constrained quadratic programming} (BoxQP) problem~\cite{angelis1997quadratic}. This is the simplest non-convex optimization
problem one can attempt to solve, yet it is NP-hard~\cite{murty1985some} and no
reasonably efficient classical algorithms for solving it exist~\cite{burer2009globally, bonami2018globally, vries2022tight}.\footnote{Our simulators of these CCVMs are publicly available as an open source Python package: \url{https://github.com/1QB-Information-Technologies/ccvm/}}

This paper is organized as follows. In~\cref{sec:sdes}, we introduce the
various SDEs used to solve the BoxQP problem
in our benchmarking study, including the systems of SDEs that represent the
dynamics of the DL-CCVM and the MF-CCVM. In~\cref{sec:solvingbyCIM}, we propose methods by which
these algorithms for solving BoxQP problems can be experimentally realized. Our
benchmarking results are presented in~\cref{sec:results}. In~\cref{sec:time-evolution}, we analyze the time evolution of the optical pulses in the DL-CCVM scheme when solving BoxQP problems. We conclude with a
brief discussion and prospects for future research in \cref{sec:conclusion}.

\section{\label{sec:sdes} SDE\lowercase{s} Governing CCVM Dynamics}

Let $f\!\!: \mathbb R^N \to \mathbb R$ be a real-valued differentiable function of
$N$ variables. To explore the dynamics of the proposed CCVM as a heuristic for the
global optimization of $f$, we consider six systems of SDEs. The first system of SDEs is
typical \textit{Langevin dynamics}, the convergence properties of which have been well-studied in mathematics and computer science~\cite{roberts1996exponential}:
\begin{equation}
d c_i = -\partial_i f(c)\, dt + \sigma d W_i
\quad \forall i \in \{1, \ldots, N\}.
\label{eq:sde_langevin}
\end{equation}
Here, $c_i$ are $N$ continuous random variables, and
\mbox{$d W_i\sim \sqrt{dt} N(0, 1)$,}
hence injecting Gaussian noise of with a mean of zero 
and variance of $dt$ into the dynamics. We use $\partial_i$
to denote the $i$-th partial derivative $\partial_i= \partial/\partial c_i$. For example, if
$f(c)= \sum_{i, j=1, \ldots, N} \xi_{ij} c_i c_j$ is a quadratic function,
then the $i$-th drift term is \mbox{$-\partial_i f(c)= -\sum_{j=1}^N \xi_{ij} c_j$,}
which may be realized physically via two-mode interactions.

The second system of SDEs is a modification of Langevin dynamics of the form
\begin{equation}
dc_i = (-1 + p - c_i^2)c_i\, dt - \partial_i f(c)\, dt + \sigma d W_i.
\label{eq:pumped_langevin}
\end{equation}
Here, the equation of motion  [\cref{eq:sde_langevin}] is augmented with an
additional drift term representing three physical processes: the constant $-1$
represents a photon loss rate that is normalized to that of the optical
cavity; the parameter $p$ represents the strength of the external pump field
(e.g., a laser), normalized to the photon decay rate at the signal frequency $\gamma_s$,
inducing an amplification of field strengths in the cavity during the process;
and the term $c_i^2$ results from  the nonlinear self-interaction induced by a
nonlinear crystal (e.g., a periodically poled lithium niobate
crystal (PPLN))~\cite{wang2013coherent}. We refer to this system of SDEs as \textit{pumped
Langevin dynamics}.

The third system of SDEs is akin to the continuous-time model for the optical DL-CIM~\cite{takata2015quantum, maruo2016truncated}, a fully quantum model for
describing the trajectories of the amplitudes of the DOPO pulses inside the
cavity via the positive \mbox{P-representation} expressed in terms of pairs of the
random variables $c_i$ and $s_i$ representing the in-phase and quadrature-phase components of the signal field~\cite{Marandi2014}
\begin{equation}
\begin{split}
dc_i = & \left[\left(-1+p - c_i^2 - s_i^2\right)c_i - \partial_i f(c) \right]dt
\\ & + \frac{r(t)}{A_s}\sqrt{c_i^2 + s_i^2 + \frac{1}{2}}\; dW_{i1} , \\
ds_i = & \left[\left(-1-p - c_i^2 - s_i^2\right)s_i - \partial_i f(s) \right]dt
\\ & + \frac{1}{r(t)A_s}\sqrt{c_i^2 + s_i^2 + \frac{1}{2}}\; dW_{i2} ,\\
\end{split}
\label{eq:sde_DL-CIM}
\end{equation}
where the Wiener increments $dW_{i1}$ and $dW_{i2}$ are  independently
sampled from an identical distribution. Here, $A_s =(\gamma_p \gamma_s/2\kappa^2)^{1/2}$, with $\gamma_s$ and
$\gamma_p$ being the signal and pump decay rates, respectively. The parameter $\kappa$
represents the parameteric gain due to the second-order susceptibility of the nonlinear
crystal. The parameter $r(t)$ is a factor for controlling the variance of the noise during the time evolution. It can be generated by continually injecting a squeezed vacuum state into the open port of the beamsplitter that out-couples the optical pulses from the cavity into the delay-line network~\cite{mcmahon2016fully}.

We note that the drift term in~\cref{eq:sde_DL-CIM} may not be realizable for
arbitrary differentiable functions $f$. Even in the case of polynomials, this requires
the experimental realization of higher-order optical interactions, which is
a challenge being tackled in current  research~\cite{yanagimoto2020engineering}. Therefore, we restrict
our study to the case of quadratic functions, as two-mode interactions of
programmable strength can be realized using optical delay lines and
beamsplitters. With this in mind, we refer to the system of SDEs~\eqref{eq:sde_DL-CIM} as 
{\textit{DL-CCVM dynamics}}.

The final system of SDEs studied  follows the continuous-time model of
the MF-CIM~\cite{leleu2019destabilization, kako2020coherent}.
For the conventional MF-CIM, under the assumption that the DOPO pulses maintain a Gaussian distribution
throughout the time evolution, their mean values and variances, $\mu_i$ and
$\sigma_i$, follow the SDEs~\cite{shoji2017quantum}
\begin{equation}
\begin{split}
d\mu_i = & \left[-(1+j) + p - g^2 \mu_i^2\right] \mu_i\, dt\\
& - \lambda \partial_i f(\tilde\mu)\, dt
+ \sqrt{j}(\sigma_i - 1/2) dW_i ,\\
d\sigma_i = & 2\left[ -(1+j) + p - 3g^2 \mu_i ^2  \right]\sigma_i\, dt  \\
& - 2j(\sigma_i - 1/2)^2\, dt + \left[(1+j) + 2g^2 \mu_i ^2 \right] dt.
\end{split}
\label{eq:sde_MF-CIM}
\end{equation}
Here, $j$ is the normalized
continuous measurement strength, $\lambda$ is a hyperparameter, and $g$ is the normalized second-order
nonlinearity coefficient. We refer to this system of SDEs as {\textit{MF-CCVM dynamics}}. 
Note that, unlike in the previous systems of SDEs, the gradient
field of $f$ in the MF-CCVM, defined by~\cref{eq:sde_MF-CIM}, is evaluated at a measured mean-field amplitude vector $\tilde\mu$
which differs from the instantaneous mean-field amplitude within the cavity by
a random vector representative of the uncertainty of the continuous quantum
measurements:
\begin{equation}
\tilde\mu= \mu + \sqrt{\frac{1}{4j}}\, \frac{dW}{dt}.
\end{equation}
It is also important to note that the MF-CCVM scheme can be realized for arbitrary
differentiable functions $f$, since the feedback term $-\lambda \partial_i f
(\tilde \mu)$ is calculated via a digital processor, for example, an FPGA, GPU, an application-specific
integrated circuit (ASIC), or perhaps via on-chip photonics.

\begin{figure}[!t]
\includegraphics[width=1.0\linewidth]{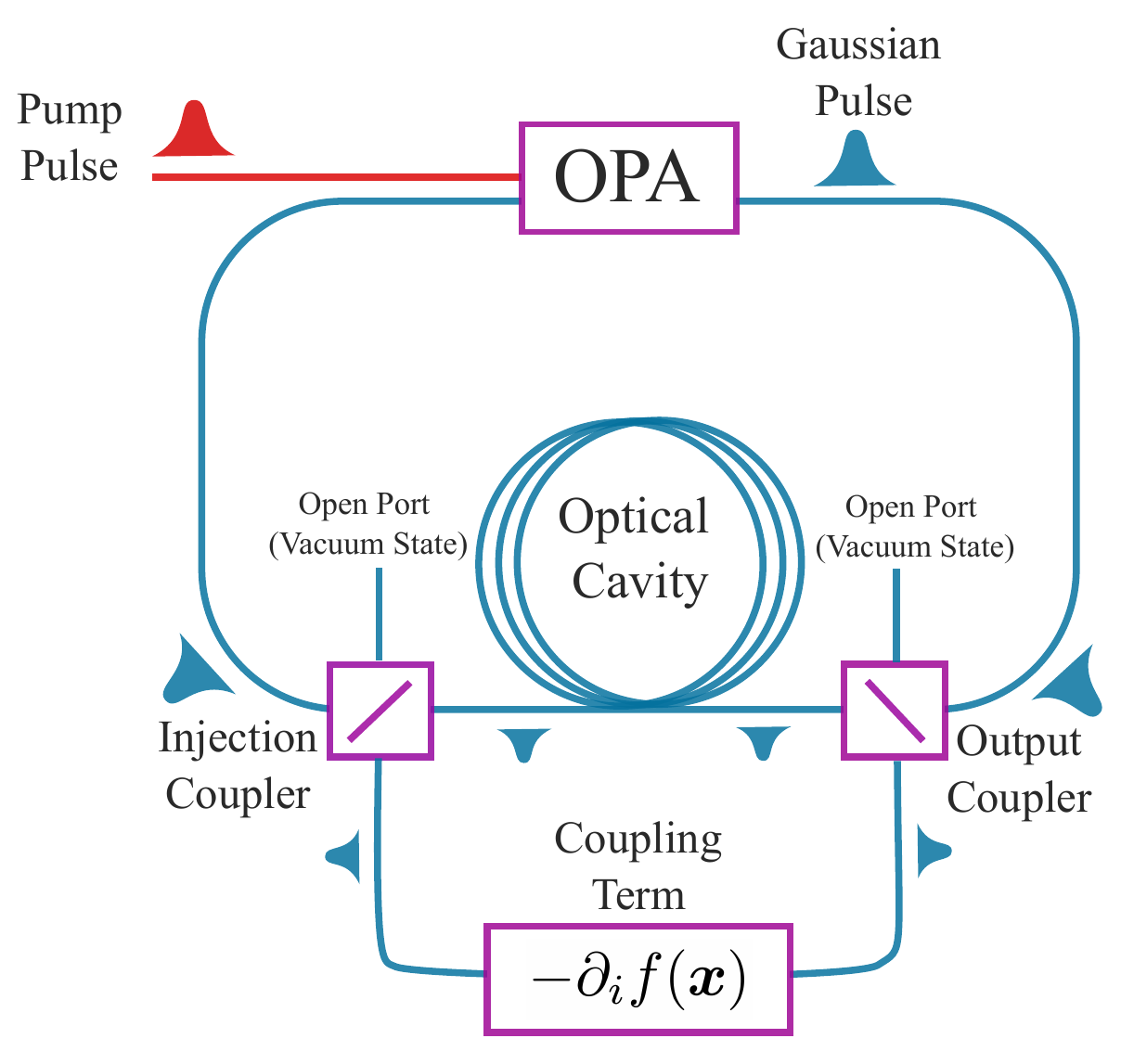}
\caption{Schematic of the proposed architecture of a CCVM. Squeezed coherent states of light are generated and amplified using an OPA element that inclues a nonlinear crystal pumped by a laser that, together with a ring cavity, constitutes a DOPO. Information is encoded in time-multiplexed oscillations of the resonator, which are coherently amplified each time they pass through the OPA element. The coupling term is implemented by taking a portion of each pulse using an output coupler, calculating the gradient of the objective function using either a delay-line or measurement-feedback scheme, and feeding the result into each optical pulse using an injection coupler. These couplers, in general, can be variable beamsplitters.}
\label{fig:ccvm-schematic}
\end{figure}

\Cref{fig:ccvm-schematic} shows the general experimental scheme used to implement the DL-CIMs and MF-CIMs. These implementations are based on the conventional CIM architecture~\cite{yamamoto2017coherent}. However, the device is operated below the saturation threshold and may employ more-general mechanisms to generate the coupling term.  The machines associated with these implementations use squeezed coherent states of light in solving non-convex continuous optimization problems. By using the quadrature-field properties of the optical pulses to represent the variables of a given problem, and by controlling the noise injected into the system, a stochastic process resembling  Langevin dynamics is implemented, which typically mixes into a low-energy state of the objective function after a sufficiently long evolution period. The key element of the architecture is a DOPO, which involves an optical resonator in the form of a ring cavity, along with a nonlinear optical crystal pumped by
a laser, for processing information that is encoded in time-multiplexed oscillations of the resonator. Using the process of optical parametric amplification (OPA), which is based on spontaneous parametric down-conversion, squeezed states are generated and continually amplified. The associated time-multiplexed pulses are coherently and iteratively coupled with the gradient of the objective function, forming the key part of the drift term of the dynamics. The gradient can be implemented by either a delay-line or measurement-feedback scheme. The diffusion process is realized by the injection of quantum noise during the OPA process, as well as the quantum measurement process in the case of the MF-CCVM.

Whereas in our study Langevin and pumped Langevin dynamics serve as heuristic algorithms to
be implemented on standard classical hardware via Monte Carlo simulations,
the DL-CCVM and the MF-CCVM are investigated as solvers that can
be implemented on both standard digital hardware and analog quantum-optical
devices. In our simulations, the success probabilities of these solvers are
used to predict the wall-clock TTS of the respective physical machines. The
wall-clock TTS is computed from the round-trip time of the optical pulses in the
cavity, the pulse durations, and the number of pulses; for more details,
see~\cref{sec:results}.

\section{\label{sec:solvingbyCIM} Solving B\lowercase{ox}QP Problems}

The BoxQP problem can be formulated as follows:
\begin{equation}
\begin{split}
\text{maximize} \quad& f(x) = \frac{1}{2} \sum_{i,j=1}^N Q_{ij} x_i x_j + \sum_{i=1}^N V_i x_i  ,\\
\text{subject to}\quad & \ell_i \le x_i \le u_i \quad \forall i \in \{1, \ldots, N\},
\end{split}
\label{eq:boxQP}
\end{equation}
where $Q\in\mathbb R^{N \times N}$ is a symmetric matrix, $V\in\mathbb{R}^N$ is
a real $N$-dimensional vector, and the lower and upper bounds
$\ell_i\in\mathbb{R}$ and $u_i\in\mathbb{R}$ specify the box constraints.

In both the DL-CIM and MF-CIM, a nonlinear 
crystal placed in an optical ring cavity fed by a sequence of pump laser pulses
plays a key role. A PPLN crystal is
commonly used to generate a degenerate nonlinear optical parametric process:
some pump photons with a  frequency of $\omega_p$ are down-converted to two degenerate
signal photons that have the frequency $\omega_s= \omega_p / 2$. These signal
pulses then propagate inside the optical ring cavity, and are coherently
amplified in each round trip when passing through the PPLN crystal.
We note that, in the \mbox{DL-CCVM,} the
nonlinear crystal is the only active element in
the system and amplifies the vacuum states toward coherent states representative
of the solution.

In CIMs, the pump rate of DOPOs is increased from below a bifurcation threshold to
a pump rate above the threshold. This creates a transition from squeezed vacuum states below the
threshold to coherent states with a binary phase oscillation ($0$ or $\pi$).
Meanwhile, the pulse amplitudes saturate to extreme minimum and maximum values that depend on the properties of the nonlinear crystal and the pump field
strength~\cite{wang2013coherent}. Operating in this regime allows a CIM to solve binary optimization problems. In contrast, we are interested in performing
the same experiment below the saturation threshold, where instead of a binary spin, we obtain an analog degree of freedom encoded in the linear
superposition of quadrature-amplitude eigenstates forming the squeezed coherent
state.

During each round trip, a portion of each optical pulse is coupled out of the
optical ring cavity to calculate and re-inject  the drift term $-\nabla f$.
In the \mbox{MF-CCVM} scheme, each out-coupled pulse is measured using a
homodyne measurement. This measurement serves two key roles: it creates the
Brownian motion \mbox{$\sqrt{j}(\sigma_i - 1/2) dW_i$} as required for the diffusion
process and it provides the measured mean-field amplitudes $\tilde \mu$, which are then
converted to digital information and used to compute $-\nabla f(\tilde \mu)$ via
a digital processor. The result is then converted back to an analog signal.
Intensity and phase modulators clock-locked with the external pump are used to
achieve amplitudes equal to the partial derivatives $\partial_i f
(\tilde \mu)$, and to match the phase of the newly generated pulses with that
of the original pulse. The result is injected back into the main optical
ring cavity at the appropriate time for each optical pulse~\cite
{mcmahon2016fully}.

In contrast, in the DL-CCVM scheme, the drift terms
$-\partial_i f(c)= \sum_{j=1}^N \xi_{ij} c_j$
are linear functions of pulse amplitudes. As with the DL-CIM, this coupling
term may be calculated through a network of delay lines. Each pulse is split
between optical lines of different lengths using an array of beamsplitters.
The amplitude and the phase of the pulses within each delay line are
appropriately manipulated using amplitude and phase modulators, before the
pulses join back together to create a coupling term. The round-trip time for
both architectures can be estimated as the pulse repetition time multiplied by the
number of pulses~\cite{maruo2016truncated, mcmahon2016fully}.

The continuous variables of the BoxQP problem can be natively encoded into
analog pulse amplitudes. The box constraints can be implemented either by
exploiting the saturation thresholds of the pulse amplitudes in the case of both the
DL-CCVM and the MF-CCVM, described by~\cref{eq:sde_DL-CIM} and~\cref{eq:sde_MF-CIM}, respectively, or, in the case of the 
MF-CCVM, also by clipping the amplitudes during each round trip using the digital processor.
Similarly, for the Langevin and pumped Langevin dynamics, the box constraints
can be implemented by simply clamping each amplitude $c_i$ between
the respective lower and upper bounds $\ell_i$ and $u_i$ at each iteration of Monte Carlo simulation.

More specifically, in the DL-CCVM scheme, by performing the change of variables
\begin{equation}
x_i \mapsto \frac{1}{2} \left(\frac{c_i}{s} + 1 \right)(u_i - \ell_i) + \ell_i,
\label{eq:x_subs}
\end{equation}
we can enforce the box constraint for each DOPO pulse. Here, $s$ is the
saturation amplitude for all DOPO pulses, assuming that they all saturate to the
same value. Assuming the coupling strength between the pulses is small, and the quadrature amplitudes $s_i$ are negligible, the saturation amplitude is approximately equal to \mbox{$s\simeq \sqrt(1 - p_0)$,} where $p_0$ is the final pump field strength in a linear pump schedule $p(t) =\frac{t}{T} p_0$. Therefore, the drift terms in the SDEs of the
DL-CCVM described by~\cref{eq:sde_DL-CIM} are 
\begin{align}
  -\partial_i f(c)
 = & - \sum_{j=1}^{N} Q_{ij}
\frac{u_i - \ell_i}{2s}
\left[\frac{1}{2}\left(\frac{c_j}{s} + 1\right)(u_j - \ell_j)
+ \ell_j \right] \nonumber\\
& + \frac{V_i(u_i - \ell_i)}{2s},
\label{eq:DLcoupling}
\end{align}
where we have substituted the variables in~\cref{eq:boxQP} by the expression on the right-hand side 
of~\cref{eq:x_subs}  and taken the gradient of the objective function with
respect to $c_i$. This expression is linear in the pulse amplitudes $c_j$
and thus able to be implemented using the DL-CCVM. The coefficients of the pulse
amplitudes in the above equation are equal to $- Q_{ij} \frac{(u_i-\ell_i)
(u_j-\ell_j)}{4s^2} $. They can be implemented using the intensity and phase
modulators on each delay line and for each pulse index $j$. The rest of the
terms in \cref{eq:DLcoupling} amount to the constant
\mbox{$\sum_j Q_{ij}\frac{(u_i - \ell_i )(u_j+ \ell_j)}{4s}
+ \frac{V_i (u_i-\ell_i)}{2s}$},
which can be calculated in advance and added to each pulse during each round trip.
An offset value for each pulse can be introduced by an external source (e.g., a
laser) that operates at the frequency of the DOPO pulses.

In the MF-CCVM scheme, the BoxQP problem's variables may simply be
encoded as $x_i:= \tilde{\mu}_i$. However, if the lower bounds $\ell_i$ are
non-negative, which is the case for the problems studied in~\cref{sec:results}, we again use the encoding
\begin{equation}
x_i \mapsto \frac{1}{2}\left(\frac{\tilde{\mu_i}}{s} + 1\right)(u_i - \ell_i) + \ell_i
\end{equation}
to increase the success probability of the optimizer. This encoding results in
the drift term
\begin{align}
-\partial_i f(\tilde{\mu})
 = & -\sum_{k=1}^{N} Q_{ij}
\frac{u_i - \ell_i}{2s}
\left[\frac{1}{2}\left(\frac{\tilde{\mu}_k}{s} + 1\right)(u_k - \ell_k)
+ \ell_k \right] \nonumber\\
& + \frac{V_i(u_i - \ell_i)}{2s}.
\label{eq:MFcoupling}
\end{align}
Here, $s$ is a hyperparameter that represents a saturation bound. As long as \mbox{$s < \sqrt{p - (1+j)}/ g$,} it is tuned to obtain the best performance.

\section{Benchmarking Results\label{sec:results}}
We have tested the solvers introduced  in the previous sections on randomly generated problem instances. We generated our instances using the same approach as in Ref.~\cite{vandenbussche2005branch}, where the elements of the $N \times N$ symmetric matrix $Q$  and the vector $V$  are random  integers generated uniformly between $-50$ and $+50$. The generated instances are labelled $N$-$D$-$S$, where $N$ is the problem size, $D$ is the density of the nonzero entries of $Q$ and $V$, and $S$ is the seed number of the random number generator. For simplicity, all variables are assumed to be constrained to the domain $[0,1]$, that is, $\ell_i = 0$ and $u_i = 1$ for all $i\in \{1,\ldots,N\}$. The global maxima of the generated instances were found using Gurobi~9.5~\cite{gurobi}.

\begin{figure}[!b]
\subfloat[\label{fig:tts-physical-DL}]{\includegraphics[scale=0.215]{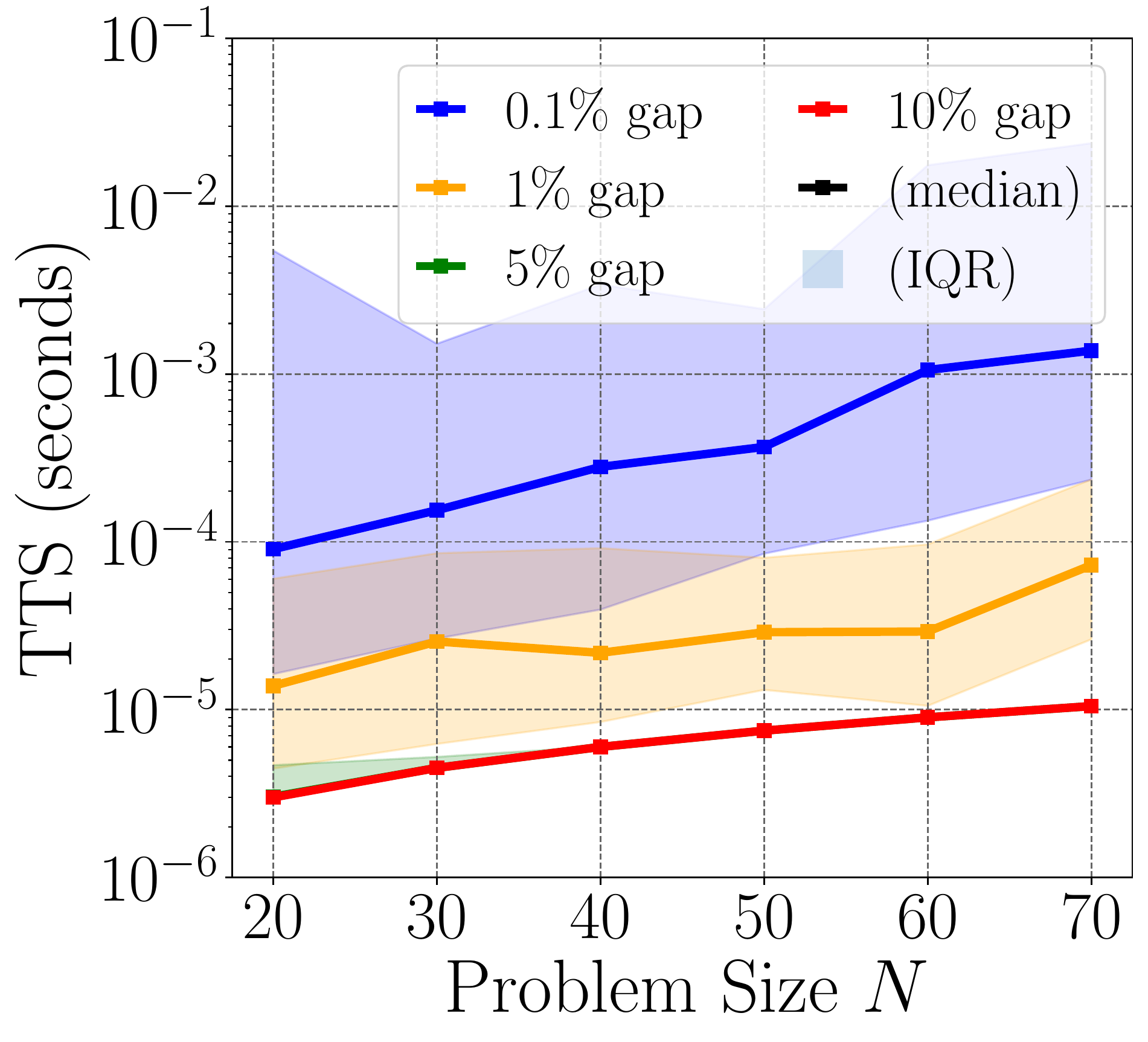}}\hspace{2mm}
\subfloat[\label{fig:tts-physical-MF}]{\includegraphics[scale=0.215]{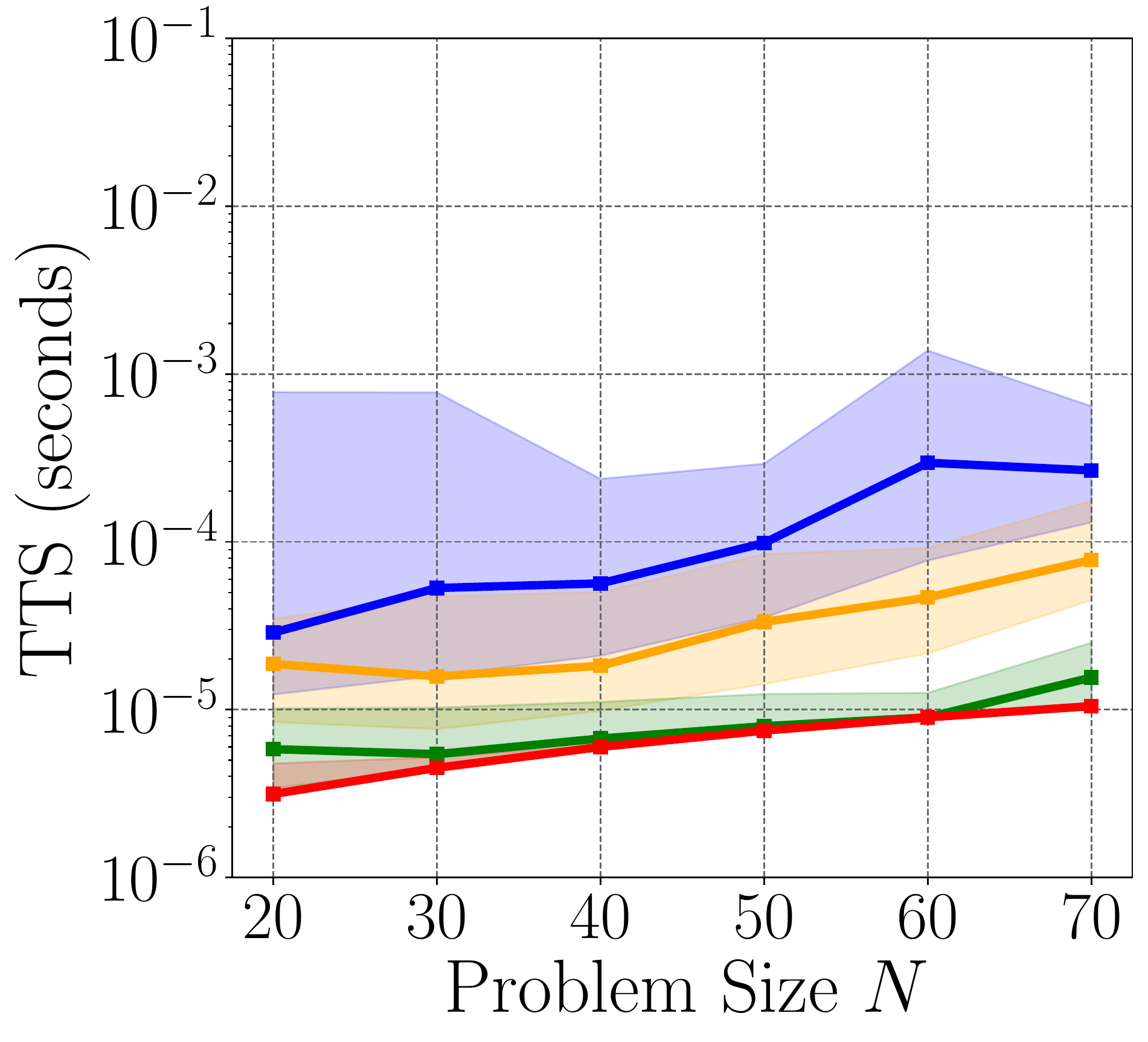}}
\caption{Estimated physical TTS for the two CCVM architectures studied: (a) DL-CCVM and (b) MF-CCVM. For each CCVM architecture, several TTS curves are displayed, corresponding to a solution within a  target percentage gap from the optimal solution. The shaded regions show the IQR corresponding to the percentile of the solved instances that were found to be within the specified gap.
\label{fig:tts-physical-MF&DL}}
\end{figure}

\begin{figure}[t]
\subfloat[\label{fig:tts-wallclock-langevin}]{\includegraphics[scale=0.215]{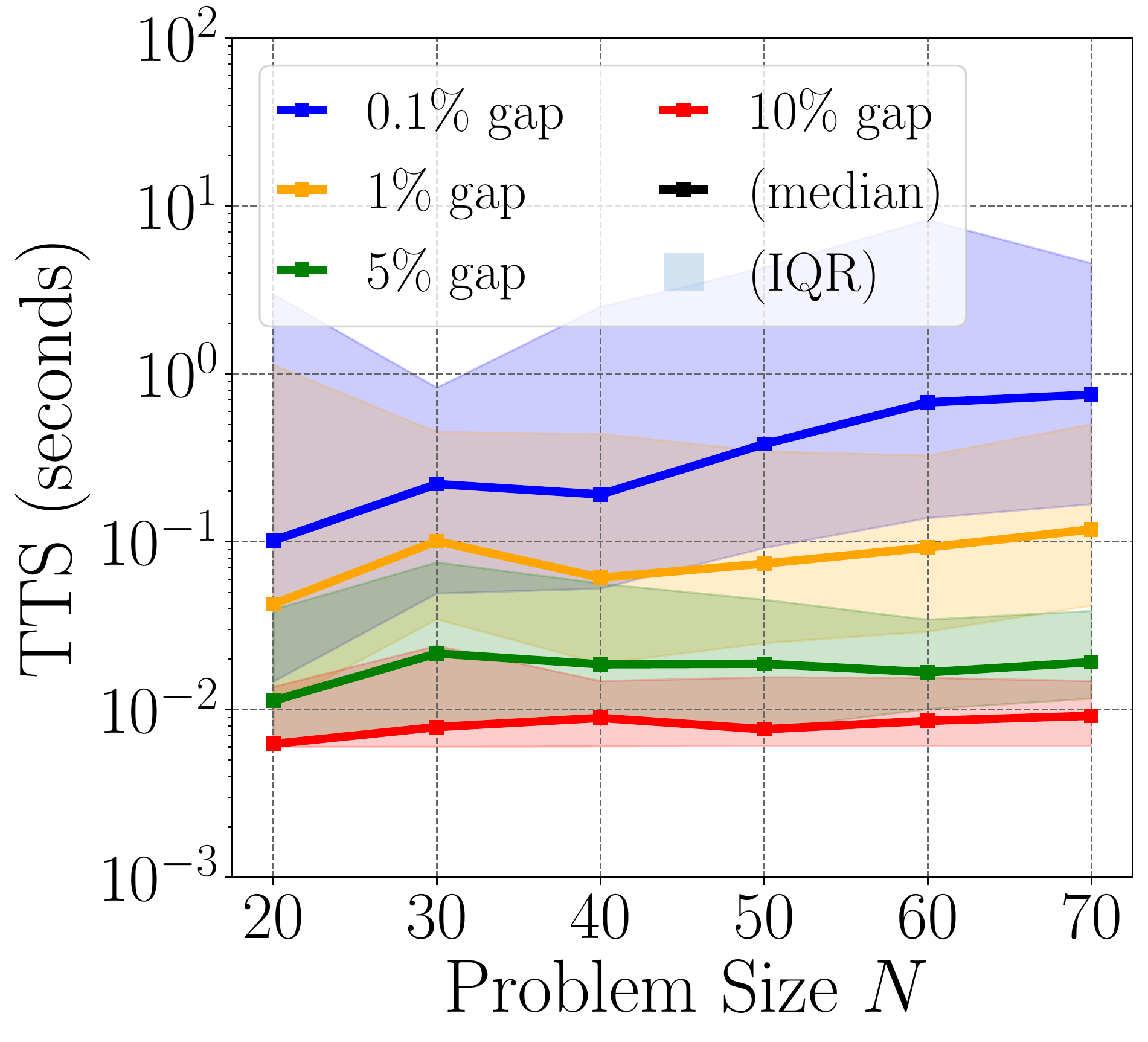}}\hspace{2mm}
\subfloat[\label{fig:tts-wallclock-pumped-langevin}]{\includegraphics[scale=0.215]{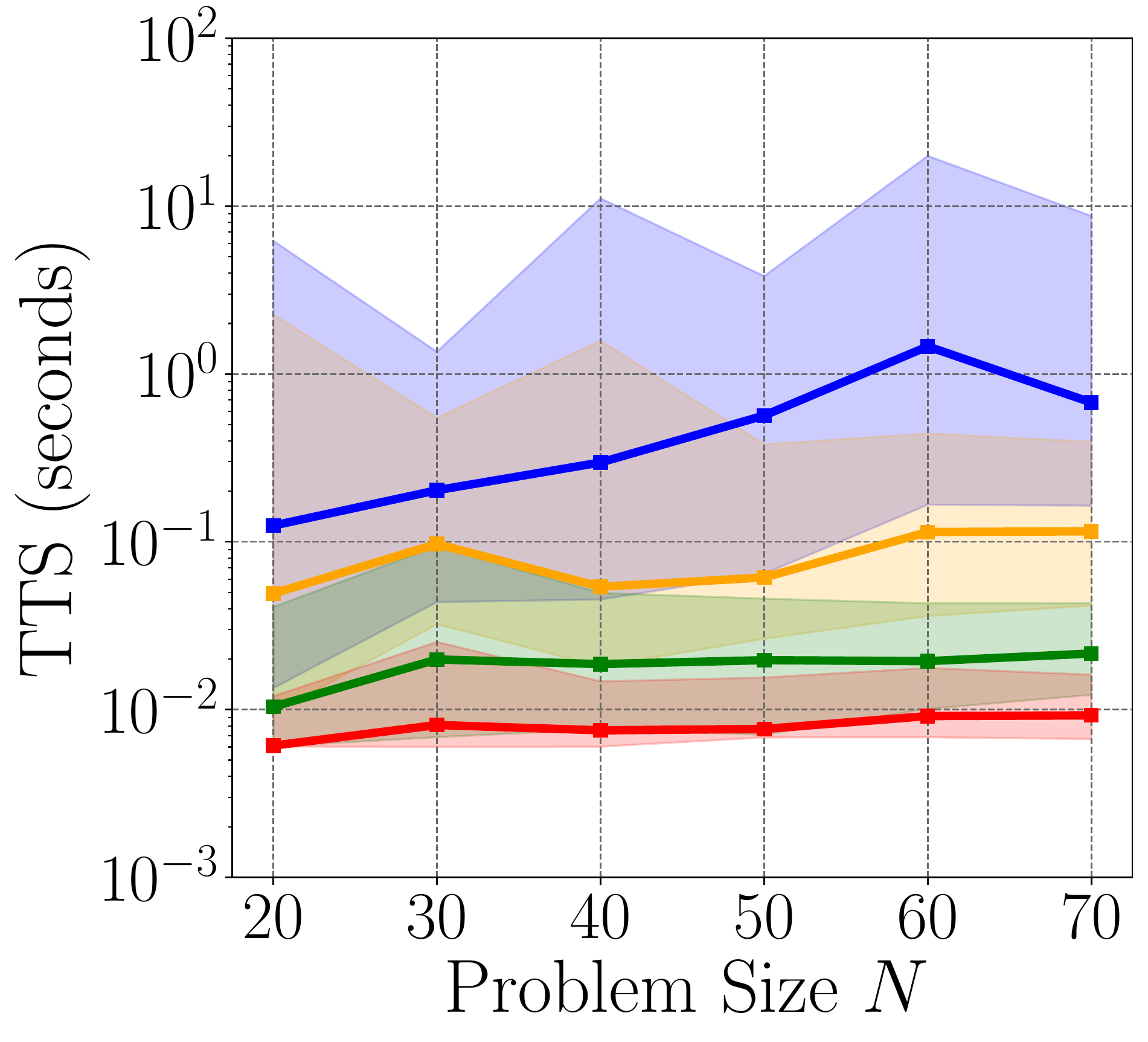}}\\ 
\subfloat[\label{fig:tts-wallclock-DL}]{\includegraphics[scale=0.215]{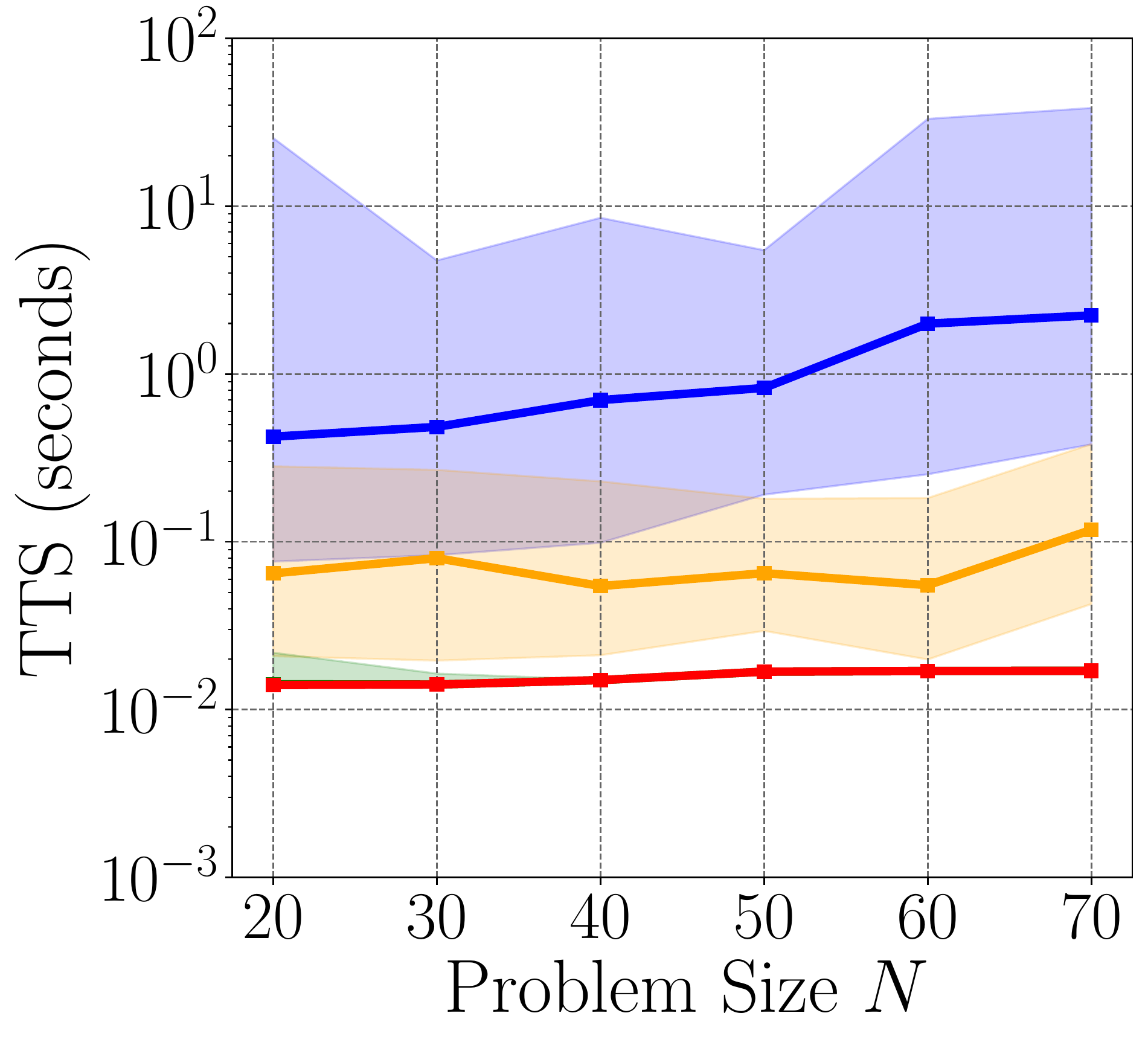}}\hspace{2mm}
\subfloat[\label{fig:tts-wallclock-MF}]{\includegraphics[scale=0.215]{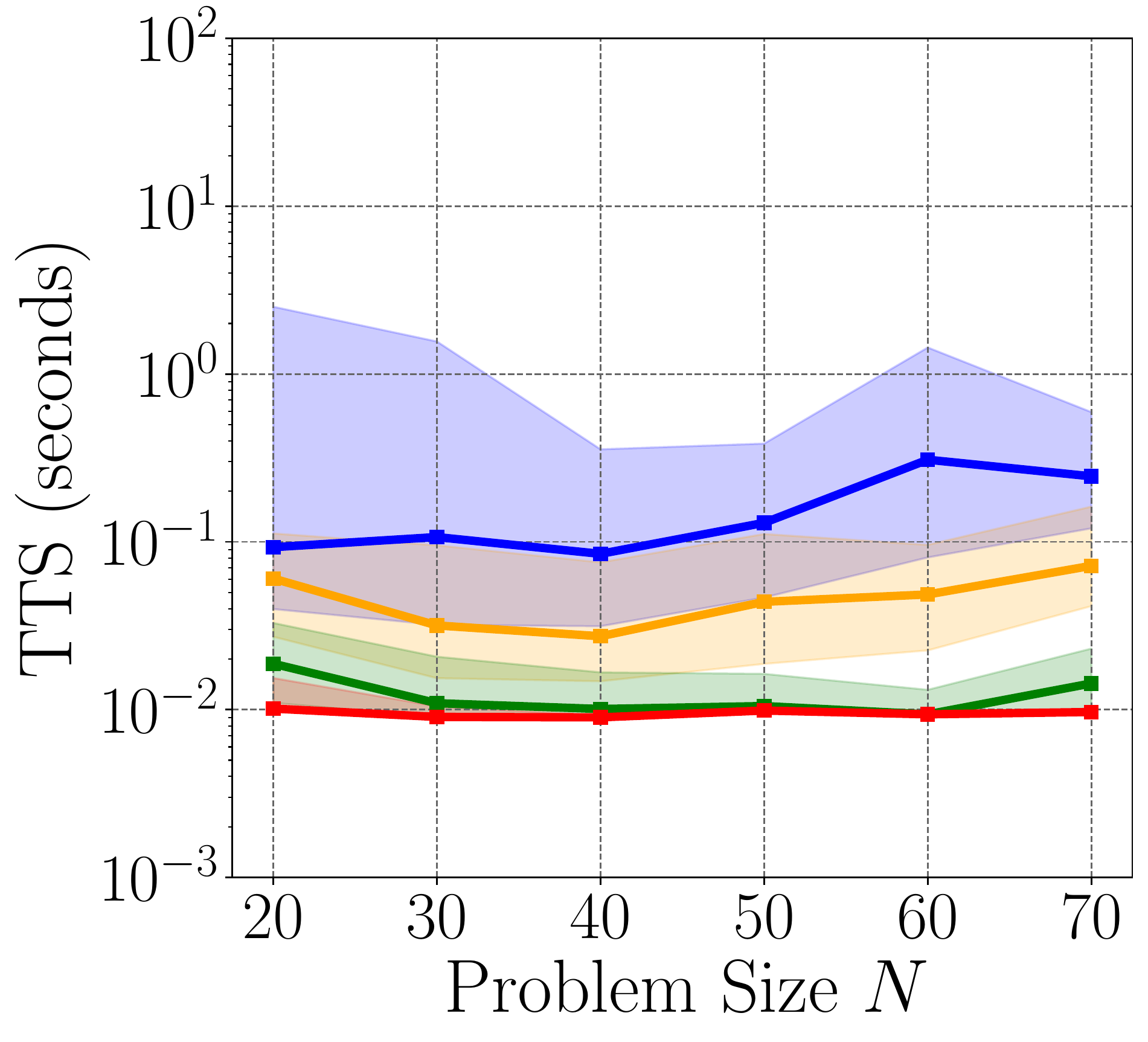}}
\caption{TTS based on the GPU time for the four solvers defined in \cref{sec:sdes}: (a) Langevin dynamics, (b) pumped Langevin dynamics, \newline (c) DL-CCVM, and  (d) MF-CCVM. For each solver, several TTS curves are displayed, corresponding to  finding a solution within a target percentage gap from the optimal solution. The shaded regions indicate the IQR for the percentage of instances that were found to be within the specified gap.
\label{fig:tts-wallclock-all}}
\end{figure}

\Cref{fig:tts-physical-MF&DL} shows the physical TTS for the two CCVM architectures introduced in~\cref{sec:sdes}. The TTS is defined as the number of  trials that must be performed to ensure a high probability (specified by some target probability of success) of observing an optimal or an acceptable approximate solution at least once, multiplied by the time required for the execution of a single trial. Here, we assume a target probability of success of 0.99. 
Thus, we compute the TTS according to the equation $\text{TTS} = R_{99} \cdot T_\text{max}$, where $R_{99} = \frac{\log(0.01)}{\log(1 - P_\text{s})}$ is the required number of trials to solve a given problem instance (i.e., observe an acceptable solution at least once) with a $99\%$ success rate, and $T_\text{max}$ is the estimated physical time required for a single trial at solving a given BoxQP instance on the physical device.
In the definition of $R_{99}$, the quantity $P_\text{s}$ denotes the success probability of a single trial; 
it is found by solving a given problem instance 1000 times and evaluating the number of times the global maximum was found for that given instance.  The time for a single trial is estimated as  $T_\text{max} = n_\text{iter} \cdot N \cdot T_\text{pulse}$, where $T_\text{pulse}$ is the time delay between two individual pulses within the cavity; for our benchmarking study, we have used the value $T_\text{pulse} = 10$ picoseconds~\cite{sankar2021benchmark}. \Cref{fig:tts-physical-MF&DL} displays the TTS for finding an approximate solution that is correct up to a multiplicative error. This is because the solutions to the problem instances are intrinsically fractional values, and due to the noise inherent to the method. \Cref{fig:tts-physical-MF&DL} shows the TTS for different multiplicative factors of the optimal value. For instance, a $0.1\%$ gap indicates that the approximate objective value found is within $0.1\%$ of the global optimal value of the problem. The  shaded regions show the interquartile range (IQR), that is, between the 25th and 75th percentiles of the solved problem instances, while the solid curves represent the median.

\Cref{fig:tts-wallclock-all} shows the wall-clock TTS (i.e., the TTS for solving the instances on a GPU) for all four solvers introduced in~\cref{sec:sdes}.
\Cref{tab:solvers_parameters} shows the parameters of the solvers used for generating \cref{fig:tts-physical-MF&DL,fig:tts-wallclock-all}. For the Langevin, pumped Langevin, and DL-CCVM solvers, the pump field $p(t) =\frac{t}{T} p_0$ follows a linear schedule, while for the MF-CCVM it is set to \mbox{$p(t) = \frac{t}{T} p_0 + 1 + j(t)$} to compensate for the measurement loss and background loss. 
On the other hand, the measurement strength $j(t) = j_0 \exp\left(-\alpha \frac{t}{T}\right)$, where $\alpha$ is an arbitrary pararmeter. In this scheme, the measurement strength is programmed according to the required amount of noise in the stochastic process. For the DL-CCVM solver, the injected noise parameter \mbox{$r(t) = r_0 \exp\left(-\beta \frac{t}{T}\right)$,} where $\beta$ is an arbitrary parameter.

\begin{table}[b]
\begin{tabular}{|c||c|c|c|c|}
\hline
\multirow{2}{*}{Parameter} & \multirow{2}{*}{Langevin} & Pumped & \multirow{2}{*}{DL-CCVM} & \multirow{2}{*}{MF-CCVM} \\
 & & Langevin & & \\ 
\hline
$n_\text{iter}$ & $15000$ & $15000$ & $15000$ & $15000$ \\
\hline
$p_0$ & -- & $1.5$--$2.0$ & $1.75$--$2.5$ & $0.1$--$1.0$ \\
\hline 
$dt$ & $0.005$--$0.01$ & $0.005$--$0.01$ & $0.005$--$0.05$ & $0.0025$ \\
\hline
$\sigma$ & $0.02$--$0.5$ & $0.02$--$0.5$ & -- & -- \\
\hline
$(r_0, \beta)$ & -- & -- &  $(10, 3)$ & -- \\
\hline
$A_s$ & -- & -- & 10 & -- \\
\hline
$s$ & -- & -- & $\sqrt{1 - p_0}$ & $0.2$ \\
\hline
$(j_0, \alpha)$ & -- & -- & -- & $(20, 3)$ \\
\hline
$\lambda$ & -- & -- & -- & $10$--$20$ \\
\hline
$g$ & -- & -- & -- & $0.01$ \\
\hline
\end{tabular}
\caption{Parameters for the Langevin, pumped Langevin, DL-CCVM, and MF-CCVM solvers.}
\label{tab:solvers_parameters}
\end{table}

\begin{figure}[!t]
  \includegraphics[width=0.75\linewidth]{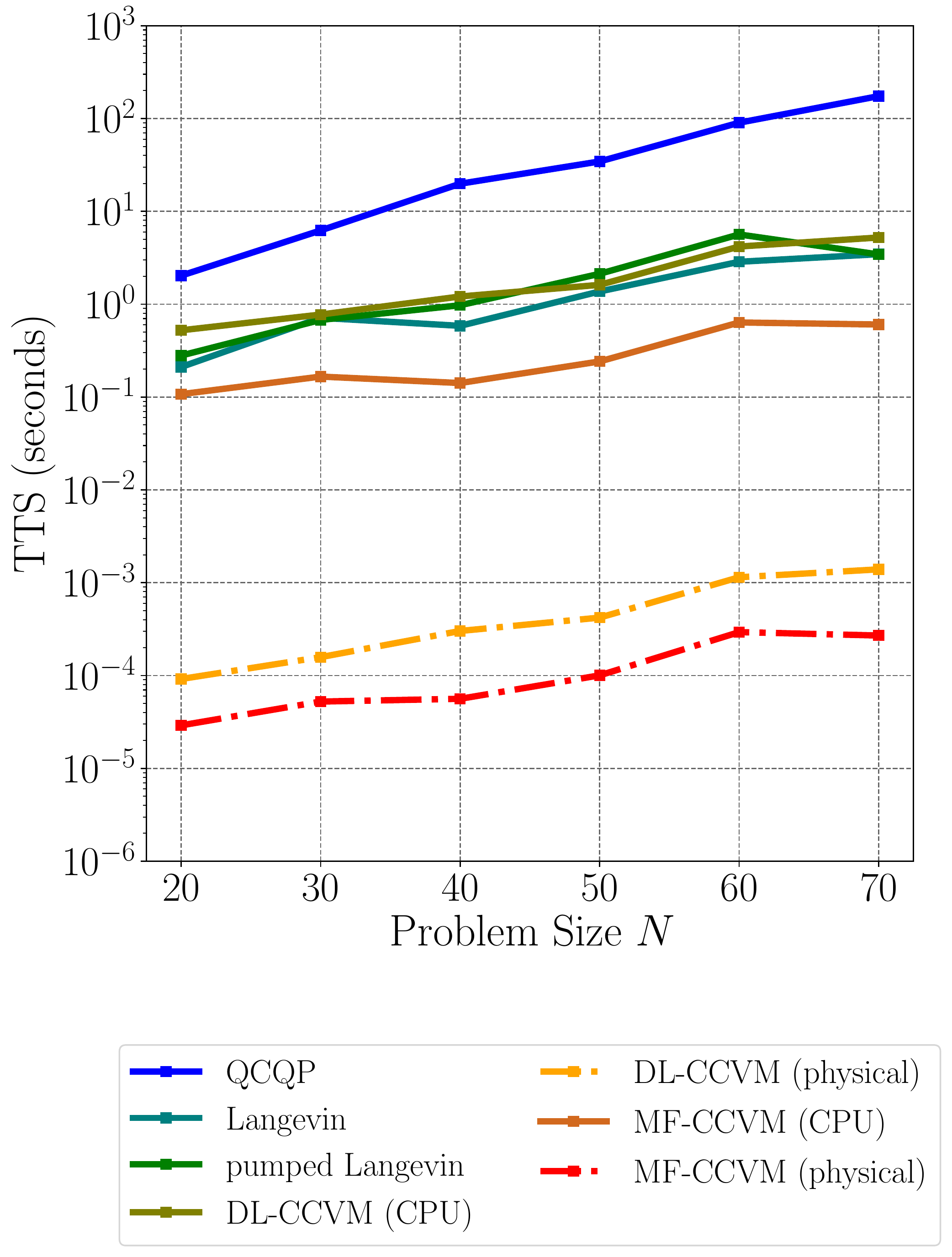}
  \caption{Wall-clock TTS for optics-based CCVM implementations (dashed--dotted curves) in comparison with the wall-clock TTS for various classical solvers implemented on conventional digital devices (solid curves). The latter have been computed based on the CPU time, while the former were calculated using the parameters of the corresponding physical device. The estimated TTS values for optics-based CCVM solvers are more than three orders of magnitude lower than the CPU wall-clock TTS values for the conventional digital solvers.
  \label{fig:tts-all-solvers}}
  \end{figure}

\Cref{fig:tts-all-solvers} shows our benchmarking comparison of the estimated physical TTS for the optics-based implementations of DL-CCVM and MF-CCVM, the TTS for all four solvers introduced in~\cref{sec:sdes} implemented on conventional digital devices (i.e., those based on electronics, not optics), and the TTS for the quadratically constrained quadratic programming (QCQP) solver~\cite{park2017general}, a general heuristic solver for quadratic programming problems. Optics-based CCVMs can be several orders of magnitude more performant than their conventional digital counterparts, and, especially in the case of the DL-CCVM, be much more power efficient.

  \begin{figure}[thb]
  \subfloat[\label{}]{\includegraphics[scale=0.215]{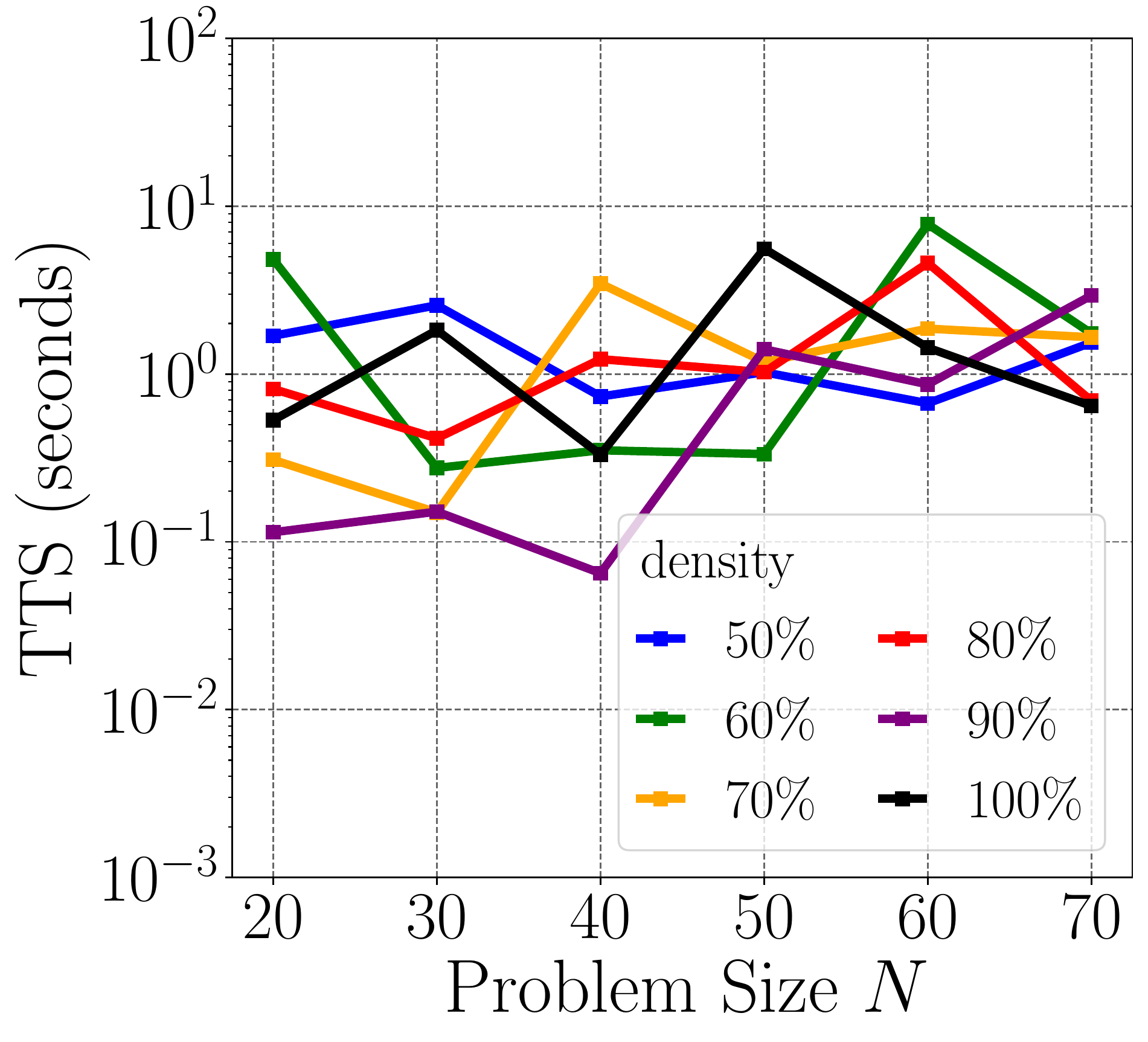}}\hspace{2.0mm}
  \subfloat[\label{}]{\includegraphics[scale=0.215]{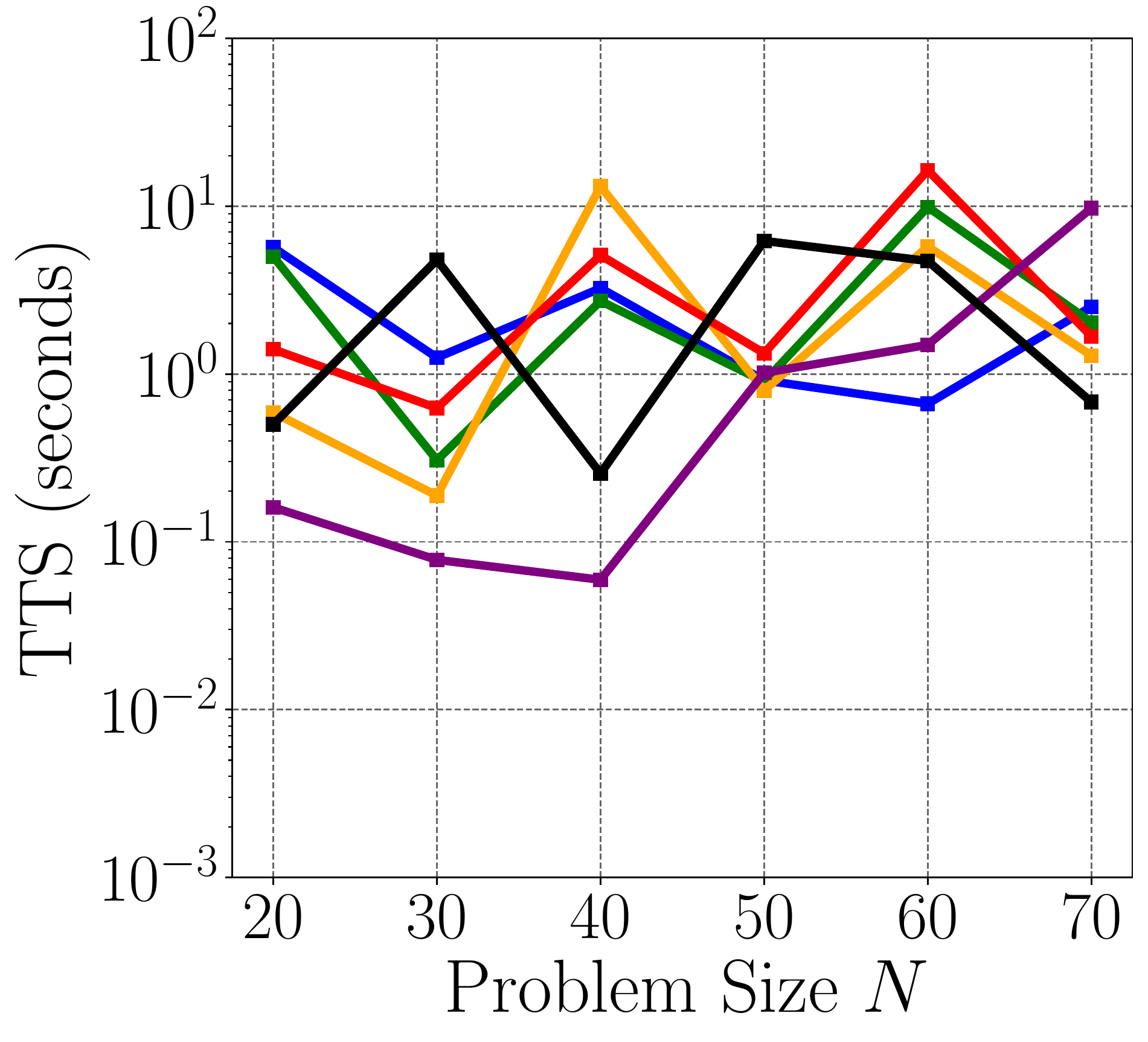}}\\
  \subfloat[\label{}]{\includegraphics[scale=0.215]{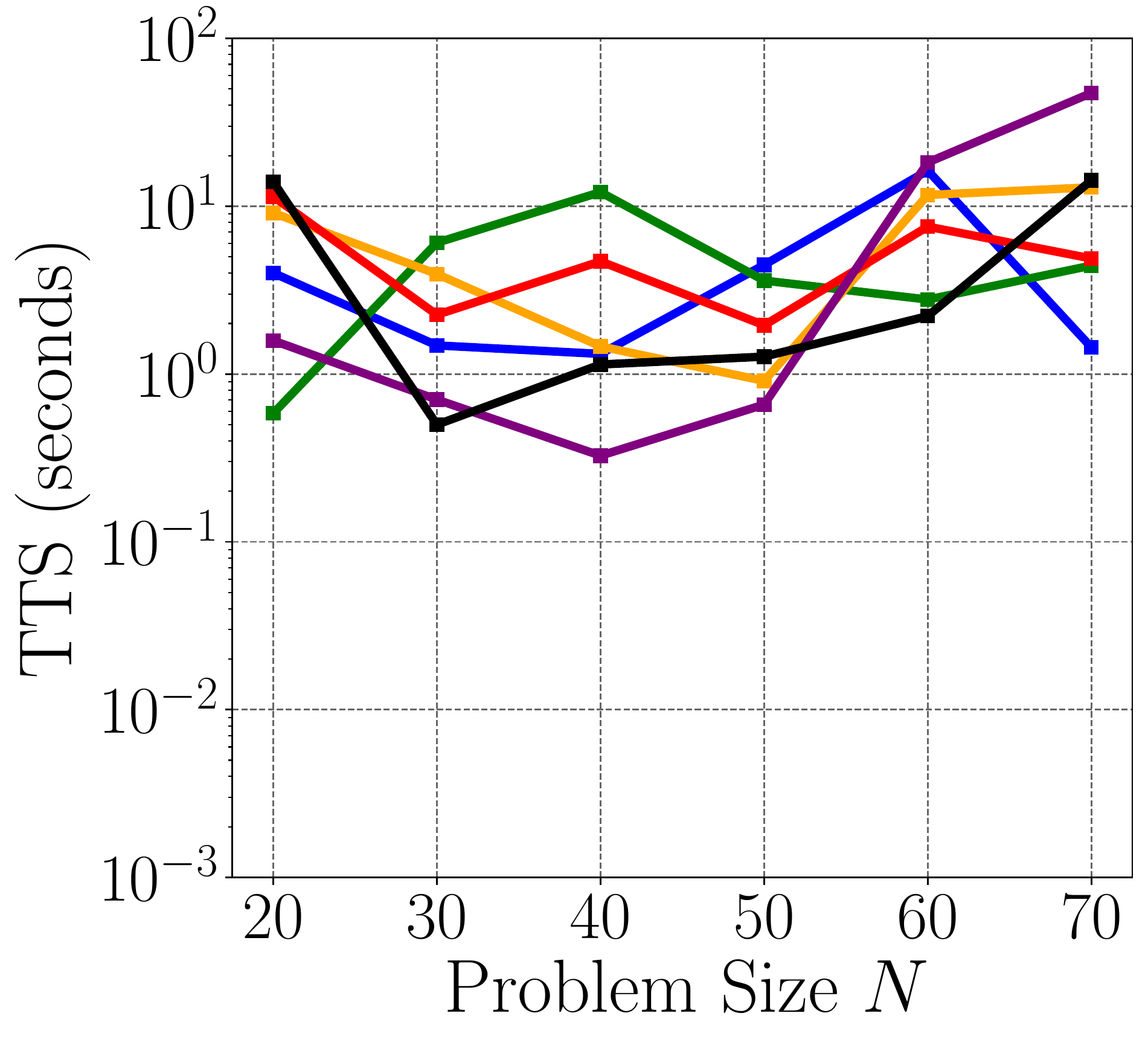}} \hspace{2.0mm}
  \subfloat[\label{}]{\includegraphics[scale=0.215]{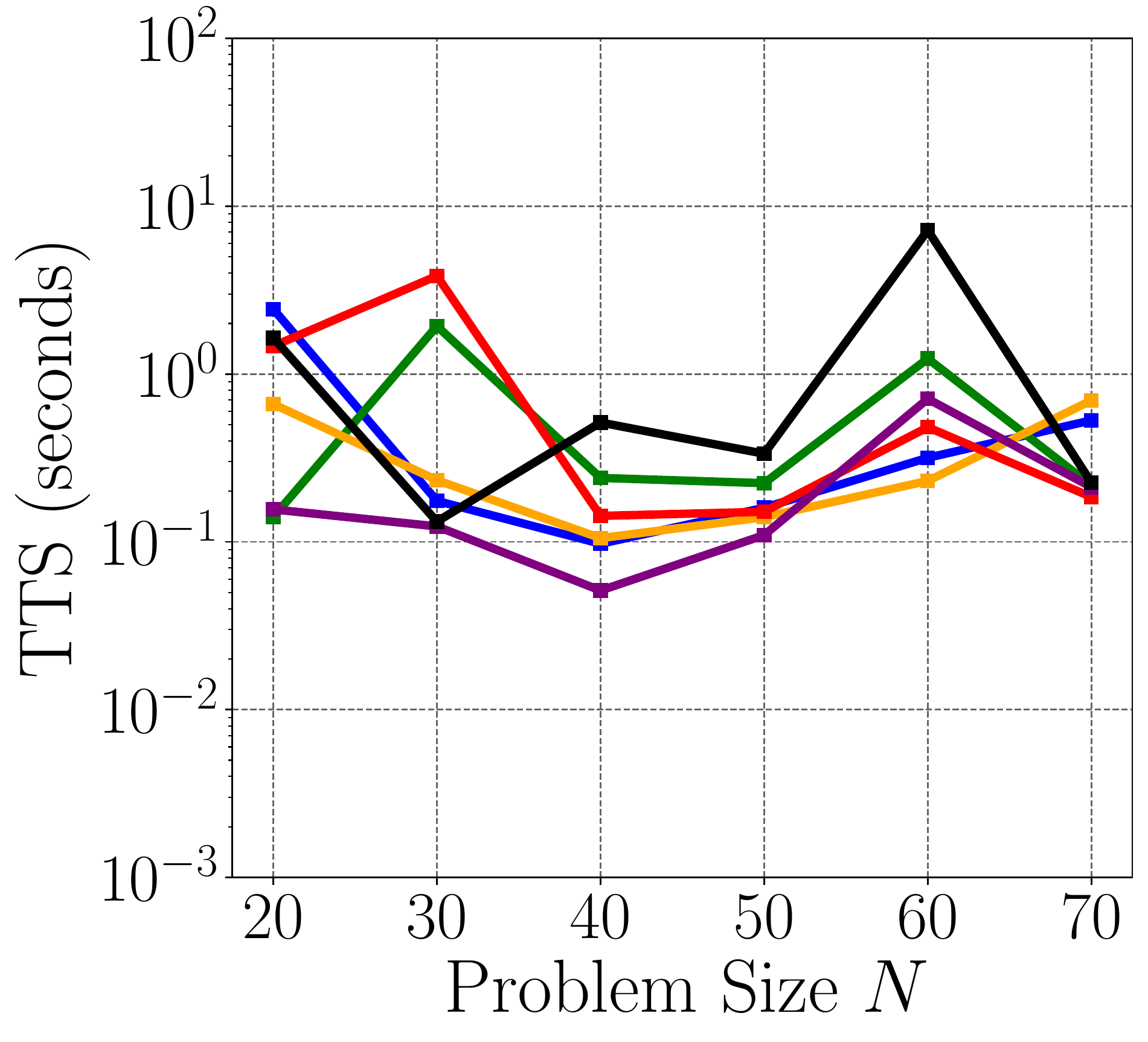}}
  \caption{TTS based on the GPU time for the four solvers defined in~\cref{sec:sdes}, for different densities of the matrix $Q$ and the vector $V$: \newline (a) Langevin dynamics, (b) pumped Langevin dynamics, (c) DL-CCVM, and (d) MF-CCVM. Solutions within a $0.1\%$ gap of the optimal value were found in all cases. For all solvers, no evident dependency on the density was observed. \label{fig:tts-density-wallclock-all}}
  \end{figure}

\Cref{fig:tts-density-wallclock-all} demonstrates a lack of dependency of the performance of the solvers on the density of  $Q$ and $V$. Specifically, it shows the wall-clock TTS of the six solvers defined in~\cref{sec:sdes} for finding a  solution within a $0.1\%$ gap from the optimal solution, individually calculated for various density values of the problem instances. While for exact solvers, such as CPLEX and Gurobi, the TTS heavily depends on the density~\cite{vandenbussche2005branch, gurobi}, for the solvers analyzed here, the TTS does not show any significant dependency on it. This is because, unlike the exact solvers, the heuristic solvers analyzed here do not exploit the sparsity of problem instances.

\section{Time-Evolution of the DL-CCVM}
\label{sec:time-evolution}

\begin{figure}[!h]
\subfloat[\label{fig:time-evol-instance1}]{\includegraphics[scale=0.215]{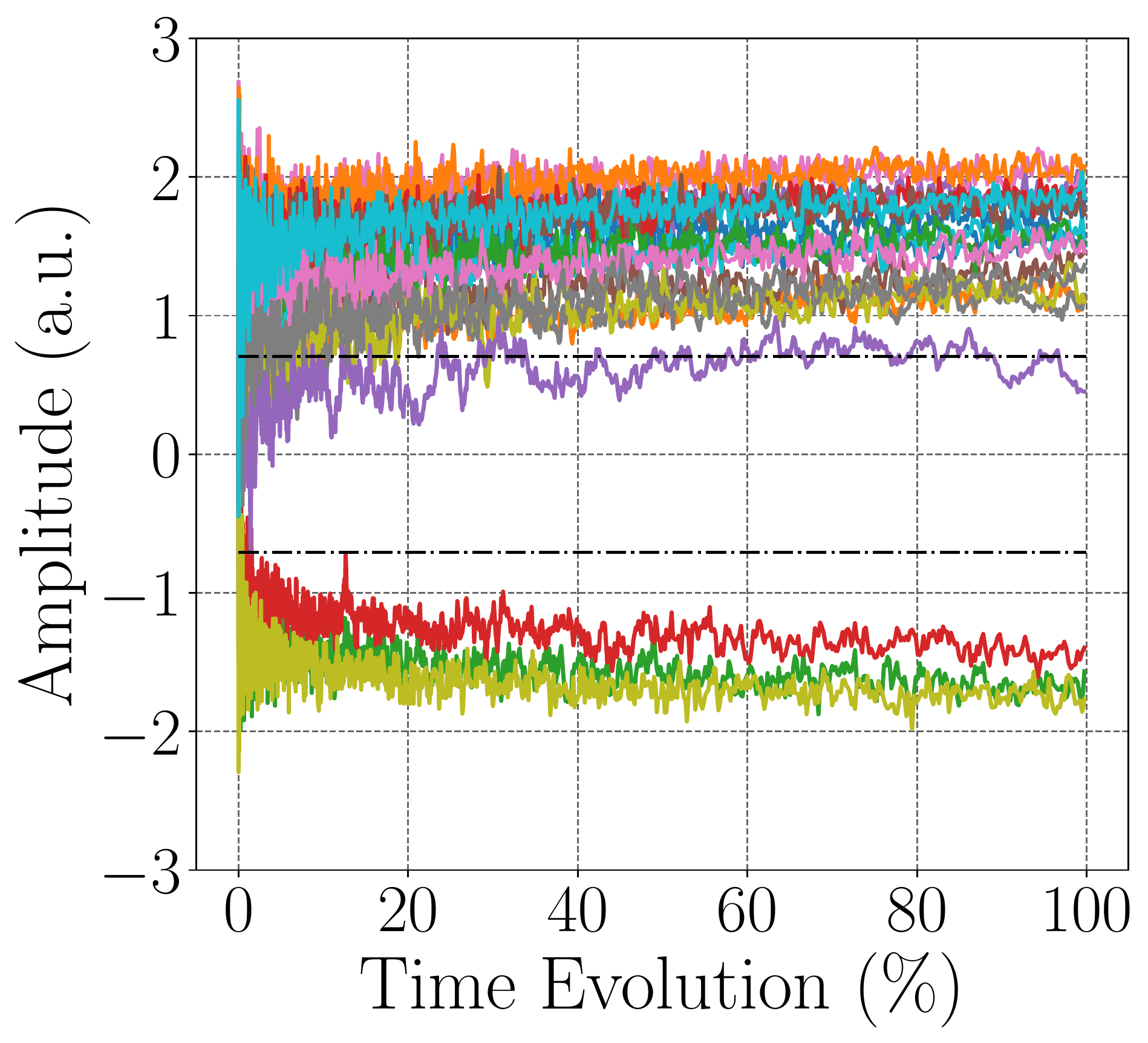}}\hspace{2mm}
\subfloat[\label{fig:time-evol-instance2}]{\includegraphics[scale=0.215]{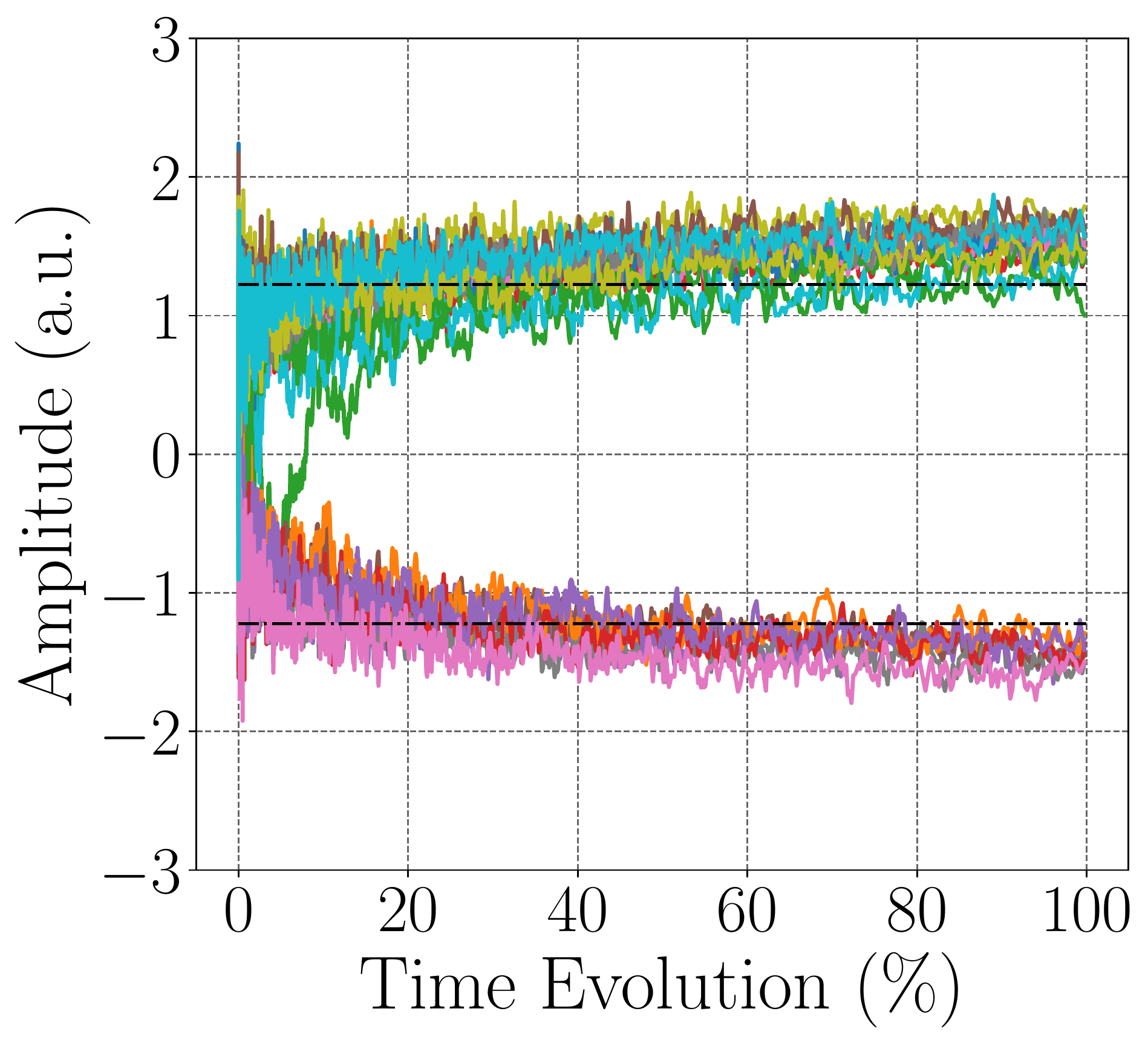}}\\
\subfloat[\label{fig:time-evol-instance3}]{\includegraphics[scale=0.215]{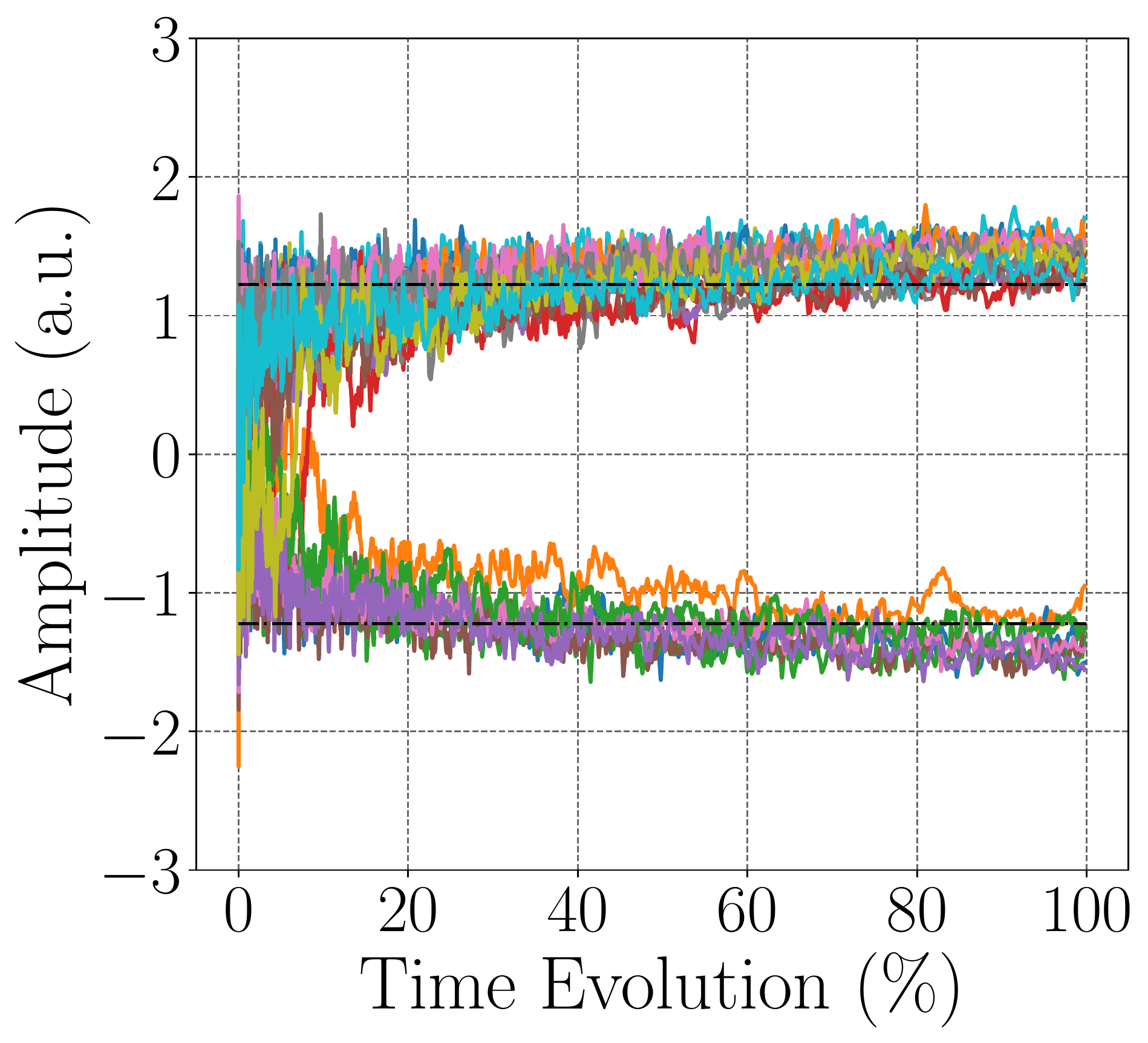}}\hspace{2mm}
\subfloat[\label{fig:time-evol-instance4}]{\includegraphics[scale=0.215]{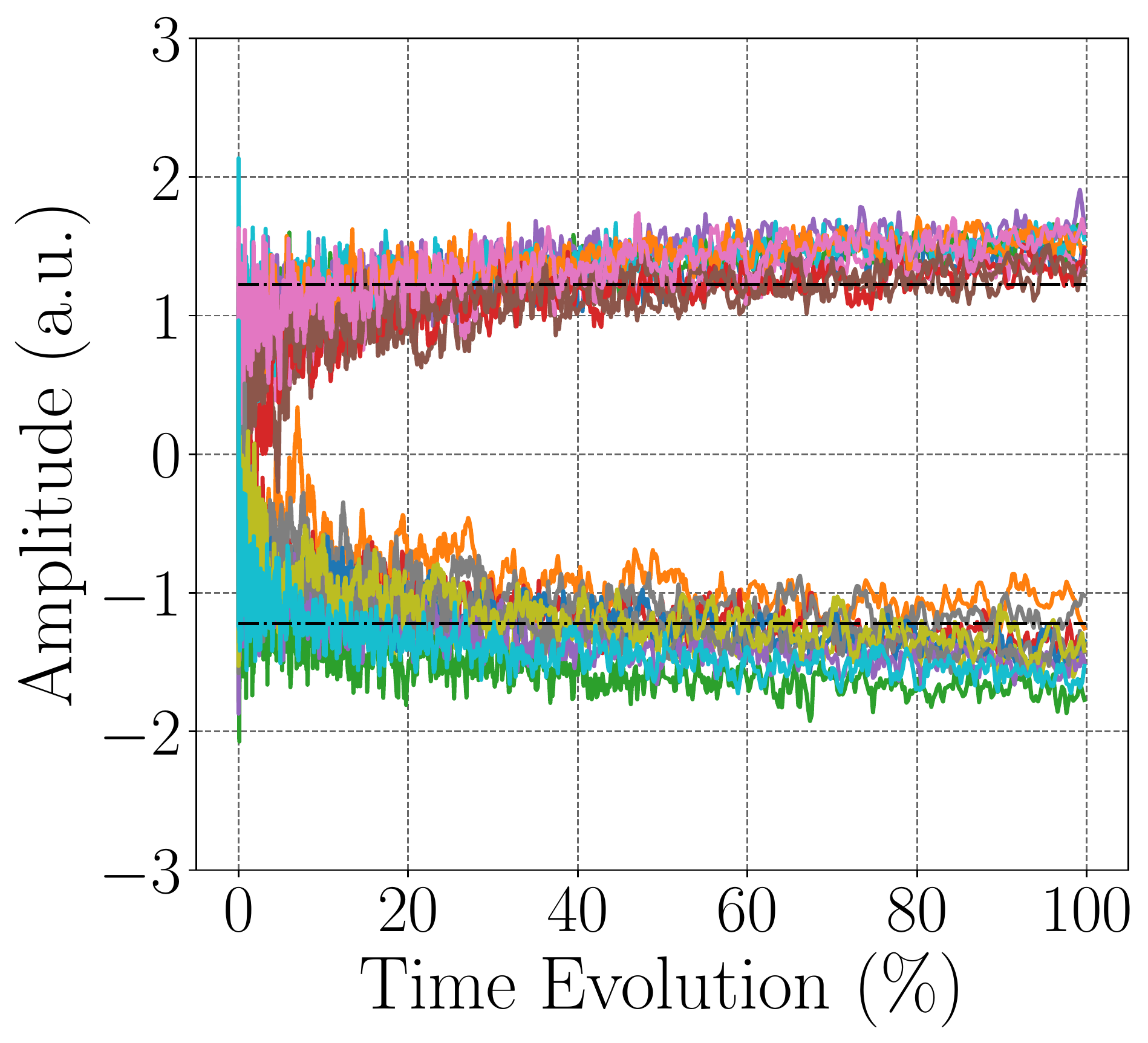}}
\caption{Time evolution of the DOPO pulses' amplitudes for four randomly generated BoxQP problem instances solved using the DL-CCVM: (a) \texttt{tuning020-50-2}, (b) \texttt{tuning020-90-8}, (c) \texttt{tuning020-100-3}, and (d) \texttt{tuning020-100-5}, where the format \texttt{tuning{\textit{N}}-{\textit{D}}-{\textit{S}}} represents a problem instance of size $N$, density $D$, and seed number $S$.  The heavy black dashed lines indicate the values $-s$ and $s$, at which the amplitudes are clamped prior to computing the objective function. The objective function is calculated after the substitution $x_i = \frac{1}{2}(c_i/s + 1)$ to satisfy the box constraint in~\cref{eq:boxQP}, with $l_i = 0$ and $u_i = 1$ for all $i\in\{1,\ldots,N\}$. Note that, for each of the problem instances, the majority of the amplitudes have an  optimal value at the boundary of the box constraint, while only some of the amplitudes converge to fractional values.}
\label{fig:time-evolution_DL-CIM}
\end{figure}

\Cref{fig:time-evolution_DL-CIM} shows the time evolution of the DOPO pulses' amplitudes for the DL-CCVM dynamics~(see \cref{eq:sde_DL-CIM}). A randomly generated problem instance of size $N=20$ was used for the results displayed in each panel. The heavy dashed lines indicate the values $-s$ and $s$ at which the amplitudes are clapmed at the end of the process. The system parameters, including the pump value, are optimized for the highest success probability through parameter tuning. Their ranges are shown in~\cref{tab:solvers_parameters}. Since the clamping is applied at the end of the evolution, it can be considered a post-processing step that is performed on a classical computer.
After the clamping of the amplitudes $c_i$, they are plugged into the equation $x_i = \frac{1}{2}(c_i/s + 1)$ to obtain the problem variables in order to implement the box constraint $0\le x_i \le 1$. It is evident that, for the solutions of the problems, the majority of the variables are found to be at the edges of the box constraint at $x_i = 0$ or $x_i = 1$. For the time evolutions shown in~\cref{fig:time-evolution_DL-CIM}, only one variable has a fractional value within the range $0<x_i<1$. The rest of the variables attain the value $0$ or $1$ in the optimal solution. It is easy to intuitively understand why most variables attain extreme values in BoxQP problems by imagining that, in two dimensions, the chance of a hyperbola's contours touching a corner of a randomly placed rectangle is much higher than that of becoming tangent to a side of that rectangle.

\section{\label{sec:conclusion} Conclusion}
 
We have proposed an approach for solving problems in the class of NP-hard box-constrained quadratic programming (BoxQP) problems using modifications to the dynamics of conventional coherent Ising machines (CIM). By exploiting  the fact that degenerate optical parametric oscillator (DOPO) pulses' amplitudes are intrinsically continuous, the continuous variables of a BoxQP problem can be directly encoded and processed as analog information. This allows us to avoid the associated large resource overheads  of digital (i.e., binary) information processing and discretization errors. Furthermore, the box constraints can be implemented by exploiting the amplitude saturation of the DOPO pulses, or simply by clamping the pulse amplitudes when using a measurement-feedback scheme. Our numerical results motivate using coherent continuous-variable machines (CCVM) as analog heuristic solvers for non-convex continuous optimization.

Our proposed CCVMs are intrinsically analog devices, that is, information is encoded in analog signals and processed by manipulating these signals continuously in time. Note that this is different from the notion of analog computing often associated with some  qubit-based quantum devices that do not rely on the circuit model of quantum computation (e.g., quantum annealers), which use a binary encoding in qubits, but process information in an analog manner, that is, by evolving the corresponding quantum states continuously in time. 

Our benchmarking study demonstrates that CCVMs can solve non-convex continuous optimization problems over three orders of magnitude faster than classical heuristic solvers. We do not expect CCVMs to offer a scaling advantage in the time to solution. Instead, our goal is to dramatically reduce the energy consumption of simulating Langevin dynamics compared to conventional digital computers.

Our study serves to stimulate research into experimental realizations of CCVMs. Implementing more-complicated constraints than box constraints is an avenue for further exploration. In addition, more-general objective functions may be implemented using CCVMs. Measurement-feedback mechanisms can be used to generate the gradient of an arbitrary differentiable objective function acting as a drift term. However, generating the gradient of such an arbitrary function is more challenging in the case of CCVM schemes based on optical delay lines, further motivating research into the experimental realization of higher-order optical interactions.

\section*{Acknowledgement}
The authors thank Yoshihisa Yamamoto, Edwin Ng, Satoshi Kako, and Sam Reifenstein for helpful discussions. We thank our editor, Marko Bucyk, for his careful review and editing of the manuscript. The authors acknowledge the financial support received through the NSF’s CIM Expeditions award (CCF-1918549). P.~R.~acknowledges the financial support of Mike and Ophelia Lazaridis, Innovation, Science and Economic Development Canada (ISED), and the Perimeter Institute for Theoretical Physics. Research at the Perimeter Institute is supported in part by the Government of Canada through ISED and by the Province of Ontario through the Ministry of Colleges and Universities.

\clearpage
\bibliography{refs}

\end{document}


\setcounter{section}{0}
\setcounter{equation}{0}
\setcounter{figure}{0}
\setcounter{table}{0}
\setcounter{page}{1}

\renewcommand{\theequation}{S\arabic{equation}}
\renewcommand{\thefigure}{S\arabic{figure}}
\renewcommand{\thetable}{S\arabic{table}}
\renewcommand{\thesection}{S\arabic{section}}
\renewcommand{\bibnumfmt}[1]{[#1]}
\renewcommand{\citenumfont}[1]{#1}

\begin{center}
\textbf{\large Supplementary Information:\\
Solving Nonconvex Box-Constrained Quadratic Programming Problems Using Coherent Ising Machines}
\end{center}